\documentclass{article}

\usepackage[final]{nips_2017}

\usepackage[utf8]{inputenc} % allow utf-8 input
\usepackage[T1]{fontenc}    % use 8-bit T1 fonts
\usepackage{hyperref}       % hyperlinks
\usepackage{url}            % simple URL typesetting
\usepackage{booktabs}       % professional-quality tables
\usepackage{amsfonts}       % blackboard math symbols
\usepackage{nicefrac}       % compact symbols for 1/2, etc.
\usepackage{microtype}      % microtypography
\usepackage{amsmath}
\usepackage{amsfonts}
\usepackage{subcaption}
\usepackage{color}
\usepackage{listings}
\usepackage{todonotes}
\usepackage{colortbl}
\usepackage[framemethod=tikz]{mdframed}
\usepackage{tikz}
\usetikzlibrary{
backgrounds,
mindmap,
shapes.geometric,
arrows,
trees
}

\newcommand{\argmin}{\mathrm{argmin}}

\usepackage[ruled]{algorithm2e} % For algorithms

\SetAlFnt{\small}
\SetAlCapFnt{\small}

\SetAlCapNameFnt{\small}
\SetAlCapHSkip{0pt}
\IncMargin{-\parindent}

\title{The Importance of Constraint Smoothness for Parameter Estimation in Computational Cognitive Modeling}

\author{
  Abraham Nunes \thanks{Corresponding author (\texttt{nunes@dal.ca})} \\ 
  Dalhousie University, Halifax, NS  \\
  \And 
  \And Alexander Rudiuk \\
  Dalhousie University, Halifax, NS
}

\begin{document}
% \nipsfinalcopy is no longer used

\maketitle
\begin{abstract}

Psychiatric neuroscience is increasingly aware of the need to define psychopathology in terms of abnormal neural computation. The central tool in this endeavour is the fitting of computational models to behavioural data. The most prominent example of this procedure is fitting reinforcement learning (RL) models to decision-making data collected from mentally ill and healthy subject populations. These models are generative models of the decision-making data themselves, and the parameters we seek to infer can be psychologically and neurobiologically meaningful. Currently, the gold standard approach to this inference procedure involves Monte-Carlo sampling, which is robust but computationally intensive---rendering additional procedures, such as cross-validation, impractical. Searching for point estimates of model parameters using optimization procedures remains a popular and interesting option. On a novel testbed simulating parameter estimation from a common RL task, we investigated the effects of smooth vs. boundary constraints on parameter estimation using interior point and deterministic direct search algorithms for optimization. Ultimately, we show that the use of boundary constraints can lead to substantial truncation effects. Our results discourage the use of boundary constraints for these applications.

\end{abstract}

\section{Introduction}
\label{s:introduction}

By virtue of its formalization of decision-making---something ubiquitously impaired in mental illness---the theory of reinforcement learning (RL) is increasingly used as a framework within which to study psychopathology. At present, psychiatric nosology is almost entirely symptom-based, and trial-and-error is used to make treatment decisions \citep{Montague2012}. This is partly due to insufficient understanding of how behavioural phenotypes are linked to underlying neural information processing. To this end, there is great interest in leveraging computational models to characterize psychiatric symptoms such that they can be linked to biological processes in a more objective fashion \citep{Montague2012, Huys2016a, Paulus2016}.

Arguably the most important tools at the disposal of computational psychiatry researchers are the procedures by which RL models can be inverted based on subjects' observed behaviours. In such a paradigm, subjects perform some psychological (usually decision-making) task. Investigators then posit formal hypotheses (i.e., RL models) describing potential mechanisms by which the subject's behavioural data were generated. For example, \citet{Huys2013} inverted a series of simple Rescorla-Wagner models \citep{Rescorla1972} using subjects' behaviours on a reward-related signal-detection task. In this task, subjects tend to learn a bias to detecting rewarding stimuli. However, not only do subjects with major depressive disorder tend not to learn this reward bias, this observation can predict persistence of the illness for up to 8 weeks \citep{Vrieze2013}. By inverting RL models upon these data, \citet{Huys2013} showed that this depression effect was due not to impaired learning, but instead due to reward insensitivity: an effect that could not be shown without the model inversion approach. Further still, the prospect of computational models guiding psychopharmacological intervention has become increasingly plausible \citep{Iglesias2016}. It is therefore important to ensure that tools for model inversion are well tested and their limitations clearly identified.

Unfortunately, RL models are many-to-one functions, and multiple unique parameterizations can yield similar behavioural outputs. For this reason, the field has largely moved away from optimization and point-estimation. Currently, Bayesian model inversion by sampling---which accounts for estimate uncertainty by returning posterior distributions over parameter settings---is the dominant approach \citep{Sharp2015}. The problem with this method is that sampling procedures are resource-intensive for all but the simplest models, thereby making critical checks such as cross-validation infeasible. Consequently, some authors continue to prefer optimization approaches \citep{Gershman2016}, which also facilitate additional model-selection procedures \citep{Rigoux2014}.

Studies in which RL models are fit to behavioural data using optimization approaches often include constraints on the parameter estimates. These typically come in the form of smooth priors (penalty functions) or bound constraints. For example, a parameter that spans the unit interval may be constrained with a Beta distribution prior (which is smooth for certain parameterizations) or bound constraints (which are nonsmooth at the boundaries 0 and 1). However, the performance of these two constraint methods has not been compared.

Using the Matlab (The MathWorks; Natick, MA) implementation of pattern search (\texttt{patternsearch} function) and an interior-point method (\texttt{fmincon}), the present study compares the accuracy of parameter estimation under smooth and non-smooth constraints. We hypothesize that the use of bound constraints---as opposed to smooth penalty functions comprised of prior probability density functions---will result in truncation effects that compromise ordinal relationships in the estimates of underlying parameters. We examine this hypothesis on a novel testbed for RL model inversion.  

\section{Background}
\label{s:background}

\subsection{The K-Context N-Armed Bandit Foraging Testbed}
\label{ss:nab-testbed}

We have implemented a testbed for assessing the performance of model fitting procedures in the context of human RL studies. The task class implemented is a K-contextual N-armed bandit, which we will outline beginning with the simplest case where $K=1$ (i.e., an N-armed bandit).

Consider walking into a casino in which there are only three slot machines (this would be a 3-armed bandit task). Each slot machine has a different probability of providing a payout, and you can only play one at a time. Taking an action corresponds to playing a given slot machine. After each play, a reward is either provided or omitted. You are afforded a total of $T$ plays, and your goal is to maximize your payout. This simple situation is essentially the N-armed bandit task.

Let us now expand the N-armed bandit into a 3-contexts 3-armed bandit task. Consider if instead of there only being three slot machines, there were three rows of slot machines with three slot machines each. A casino employee determines which row you can play in at each turn, and does so randomly. She will direct you to one of the three rows, where you play one of the machines. After your turn, she directs you to one of the three rows again, where you play, and the process repeats \textit{ad infinitum}.

What makes this task challenging is that a player does not initially know the reward associated with each slot machine. This must be learned from experience. In the simplest case, the reward probabilities associated with each slot machine remain constant, and when the player discovers the best slot machine, he or she will generally stick with that choice henceforth. In other words, after an initial period of exploration, the participant will transition to exploitation of the best option. However, it is often of interest to examine foraging behaviour, in which the task must incentivize ongoing exploration. This can be done by allowing the probability of reward associated with each slot machine to drift randomly---and independently---over trials. As such, slot machines that are initially more rewarding may later become inferior, and vice versa. This is implemented by modelling each slot machine's reward probability as a Gaussian random walk over trials \citep{Daw2006a}.

A template algorithm for simulating behaviours on the K-contextual N-armed bandit  is presented in Algorithm \ref{alg:kcnab-template}. In this procedure, the \texttt{choose$\_$action} and \texttt{learning} functions are given by whichever RL model is being simulated. This model also determines which parameters are stored in $\boldsymbol\theta$. 

\begin{algorithm}
    \SetKwData{Left}{left}\SetKwData{This}{this}\SetKwData{Up}{up}
    \SetKwFunction{Union}{Union}\SetKwFunction{FindCompress}{FindCompress}
    \SetKwInOut{Input}{Arguments}\SetKwInOut{Output}{Returns}

    \Input{
        Number of arms $N \geq 2, N \in \mathbb{Z}$     \\
        Number of contexts $K \geq 1, K \in \mathbb{Z}$     \\
        Number of trials $T \geq 1, T \in \mathbb{Z}$   \\
        Number of subjects $M \geq 1, M \in \mathbb{Z}$ \\
        Number of parameters $L \geq 1, L\in \mathbb{Z}$ \\
        Model (policy and value function form) $\mathcal M$ \\
        $T \times N \times K$ array of reward probabilities (time series), $\mathbf{P}$ \\
    }
    \Output{
        $M \times N \times T$ one-hot array of choices, $\mathbf{A}$ \\
        $M \times K \times T$ one-hot array of states, $\mathbf{S}$ \\
        $M \times T$ (binary) array of rewards received, $\mathbf{R}$ \\
    }
    $\boldsymbol\theta_m \gets$ \texttt{init$\_$params} $(L, \mathcal M), \;\forall m \in {1, \ldots, M}$ \\
    \For{$m \gets 1$ \KwTo $M$}{
        Initialize $N \times K$ array of action values $\mathbf{Q}$. \\
        \For{$t \gets 1$ \KwTo $T$}{
            $\bold s_{t}^{(m)}$ = \texttt{sample$\_$context}($\mathbf{K}$) \\
            $\bold a_{t}^{(m)}$ = \texttt{choose$\_$action}($\mathbf{Q}$, $\bold s_{t}^{(m)}$; $\boldsymbol\theta_{m,:}$) \\
            $R_{t}^{(m)}$ = \texttt{return$\_$reward}($\bold a_{t}^{(m)}$, $\bold s_{t}^{(m)}$; $\mathbf{P}_{t,:}$) \\
            $\mathbf{Q} \gets$ \texttt{learning}($\bold a_{t}^{(m)}$, $R_{t,m}$, $\bold s_{t}^{(m)}$; $\boldsymbol\theta_{m,:}$) \\
        }
    }
    \caption{Simulation template for K-contextual N-armed bandit. Note: here we assume that each subject uses the same reward probability series, although this could easily be extended to use unique reward probability time series for each subject. We also assume that \texttt{choose$\_$action} and \texttt{learning} are some arbitrary policy sampling and value function update rules, respectively.}
    \label{alg:kcnab-template}
\end{algorithm}

For instance, consider a simple Rescorla-Wagner model in which the function \texttt{choose$\_$action} samples an $N \times 1$ one-hot vector $\mathbf{a}$ from a categorical distribution with action probabilities $\boldsymbol\pi$. These probabilities are governed by the softmax function

\begin{equation}
\boldsymbol\pi = \frac{
    e^{\beta \mathbf{Q} \bold s}
}{
    \sum e^{\beta \mathbf{Q} \bold s}
},
\label{eq:softmax}
\end{equation}

\noindent where the sum is over $\mathbf{Q} \bold s$, the expected value of actions at state $\bold s$. Assume the function \texttt{learning} is governed by simple Rescorla-Wagner rule:

\begin{equation}
\mathbf{Q} \gets \mathbf{Q} + \alpha (r_t - \bold a^\top \mathbf{Q}\mathbf{s}) \mathbf{a} \bold s^\top.
\label{eq:rw}
\end{equation}

\noindent In this case, the parameter vector for the $i$th subject, $\boldsymbol\theta_i$, would consist of the inverse softmax temperature $\beta$ and the learning rate $\alpha$.

The function \texttt{return$\_$reward} accepts a $N \times K$ vector of reward probabilities, and returns a scalar reward for the subject's choice at that trial. Typically, the reward is sampled from a binomial distribution parameterized by the probability associated with the subject's choice. However, it is also possible to eliminate noise from the \texttt{return$\_$reward} function by simply returning a value equal to the reward probability at that time step. This latter case corresponds to an N-armed bandit with "Noiseless Reward." Ultimately, the simulation returns subjects' behavioural data $\mathbf{D} = \lbrace \mathbf{A}, \mathbf{S}, \mathbf{R} \rbrace$.

\subsection{Model Parameters}
\label{ss:model-params}

In the above example, the inverse softmax temperature $\beta$ takes values greater than or equal to 0. It governs the degree to which the subjects choices are consistent with their expected values. For instance, consider some trial in a 2-armed bandit task where a subject's current expected values for the options were [0.51, 0.49]. In this case, the first choice is slightly more valuable than the second. A large inverse softmax temperature would exaggerate this difference and result in the first choice being exploited despite its marginally greater value. Conversely, consider the same situation with expected values [0.99, 0.01]; in this case, one would expect almost exclusively exploitative behaviour. However, if the inverse softmax temperature is sufficiently small, the probability of the subjects selecting each choice can be roughly equal. This facilitates simulation of exploratory behaviour.

Of note, the inverse softmax temperature's effect on simulated actions saturates in practice. Often, the transition from exploratory to exploitative choices can be observed between inverse softmax temperature values of 0 and 10. In our experience, above values of 10, the observable effect on choice behaviour is diminished. 

The learning rate $\alpha$ governs the degree to which the subject's expected state-action values are updated with experience. This parameter always takes values between 0 and 1, where 0 reflects fixity of beliefs about expected state-action values, and 1 indicates complete revision of $\mathbf{Q}$ at each iteration. 

\subsection{Parameter Estimation}
\label{ss:parameter-estimation}

When human subjects perform K-contextual N-armed bandit  tasks, the experimenter collects their behavioural data (here $\mathbf{D} = \lbrace \mathbf{A}, \bold S, \mathbf{R} \rbrace$), and seeks to estimate which model and corresponding parameters were most likely to have generated those data. With the optimization approach, the objective function is as follows. For the $i$th subject, we seek the \textit{maximum a posteriori} (MAP) parameter estimates $\hat{\bold w}_i$ given the subject's trial-by-trial choices:

\begin{equation}
\hat{\bold w}_i = \underset{\bold w_i}{\argmin} \;\; -\log P(\bold A_i|\bold S_i, \bold w_i)P(\bold w_i|\boldsymbol\phi),
\label{eq:param-est-objfx}
\end{equation}

\noindent where $P(\bold w_i|\boldsymbol\phi)$ describes some prior probability distribution over $\bold w_i$. The hyperparameters of this distribution are denoted by $\boldsymbol\phi$.

The log-likelihood function for subject $i$ is given as follows, for example purposes, assuming a model with inverse softmax temperature $\beta$ and a single learning rate $\alpha$:

\begin{equation}
\log P(\bold A_i|\bold S_i, \beta, \alpha) = \sum_t \beta \mathbf{a}_t^\top \mathbf{Q}_{t-1}\bold s_t - \mathrm{logsumexp}(\beta \mathbf{Q}_{t-1}\bold s_t).
\label{eq:loglik-fx}
\end{equation}

\noindent In the above equation, the effect of the learning rate $\alpha$ cannot be directly observed. However, it influences the update of $\bold Q$ from trial to trial, and so has a substantial impact on the log-likelihood. A template for the algorithmic implementation of the negative log-likelihood function is presented in Algorithm \ref{alg:nab-nll-template}.

\begin{algorithm}
    \SetKwData{Left}{left}\SetKwData{This}{this}\SetKwData{Up}{up}
    \SetKwFunction{Union}{Union}\SetKwFunction{FindCompress}{FindCompress}
    \SetKwInOut{Input}{Arguments}\SetKwInOut{Output}{Returns}

    \Input{
        $N \times T$ one-hot array of choices for subject $i$, $\mathbf{A}_i = \{ \bold a_t^{(i)} \}_{t=1}^T$ \\
        $K \times T$ one-hot array of states for subject $i$, $\mathbf{S}_i = \{ \bold s_t^{(i)} \}_{t=1}^T$ \\
        $1 \times T$ array of rewards received for subject $i$, $\mathbf{R}_i = \{ r_t^{(i)} \}_{t=1}^T$ \\

        $1 \times K$ array of parameter estimates for subject $i$, $\bold w_i$ \\
    }
    \Output{
        Negative log-likelihood $\mathcal{L}$ \\
    }
    Extract number of time steps $T$ \\
    Extract number of arms $N$ \\
    Extract number of contexts $K$ \\
    Initialize negative log-likelihood $\mathcal{L} \gets 0$ (or to $P(\bold w|\boldsymbol\phi)$ if MAP estimation) \\

    Initialize $N \times K$ array of state-action values $\mathbf{Q}$. \\
    \For{$t \gets 1$ \KwTo $T$}{
        $\mathcal{L} \gets \mathcal{L} -  \beta \bold a_t^\top \bold Q \bold s_t + $\texttt{logsumexp}$(\beta \bold Q \bold s_t)$ \\
        $\bold Q \gets \bold Q + \alpha(r - \bold a_t^\top \bold Q \bold s_t) \bold a_t \bold s_t^\top$ \\
    }
    \caption{Template for a negative log-likelihood function (single-subject) for a $K=1$ context N-armed bandit. To simply demonstrate how this is calculated, we assume that $\bold w = (\alpha, \beta)$, which are the learning rate and inverse softmax temperatures, respectively. We also show the corresponding Rescorla-Wagner update rule for $\mathbf{Q}$ This boilerplate can be extended to other models.}
    \label{alg:nab-nll-template}
\end{algorithm}

Constraints on the parameters $\bold w$ are modelled in the prior probability distribution $P(\bold w|\boldsymbol\phi)$. Imposing bound constraints amounts to specifying a uniform distribution on parameter values lying between the lower and upper bounds, respectively. Choice of this prior ensures all final estimates will lie within the bounds, thus suitably addressing unrelaxable constraints. The downside of this approach may be that lack of smoothness at the boundaries manifests in truncation effects on the parameter estimates. This will be important when the research question emphasizes the relative order of parameter values between subjects. Such a case would arise if one is making comparisons between groups; for instance, rather than estimating the precise learning rates for two groups, it may be of greater interest to know whether one group's learning rate is systematically higher or lower than the other. Truncation effects would compromise this type of comparison since many estimates that would contain ordinal information would be forced onto a single point at the boundary. Another important feature of this form of prior is that it is infinitely strong, since no amount of data can overwhelm the bound constraints.

An alternative approach for constraining the objective function is to use some prior probability distribution whose domain of support includes the feasible region for each parameter.

For instance, let $\bold w = \lbrace \alpha, \beta \rbrace$, with $\alpha \sim$Beta($\alpha$; 1.1, 1.1) and $\beta \sim$ Gamma($\beta; 5, 1$). If we assume Beta($\cdot$) and Gamma($\cdot$) are probability density functions, our objective function $f(\mathbf{w})$ becomes

$$
f(\bold w) = - \log P(\bold A_i| \mathbf{S}_i  \beta, \alpha) - \log \mathrm{Beta}(\alpha; 1.1, 1.1) - \log \mathrm{Gamma}(\beta; 5, 1).
$$

\noindent The benefit of this method is that one may retain smoothness around the constraint boundaries, thereby preserving order relationships. One potential drawback is that some resulting estimates may lie outside of the "feasible" region (insofar as feasible is defined by the experimenter). Another potential drawback is that selection of a given prior probability density can excessively bias parameter estimates, leading to overall worse estimates in cases where the prior was misspecified. Notwithstanding, in the ensuing experiments, we hypothesize that bound constraints will result in significant truncation effects around the boundaries, and that this will not be offset by any improved accuracy of parameter estimation.

\section{Methods}
\label{s:methods}

\subsection{Experimental Design}
\label{ss:experimental-design}

The present experiment was organized according to a one-way factorial design over constraint types (bound vs. smooth). Within each factor level, we permuted the following parameters: Arms, Contexts, Trials, Reward Volatility, and Reward Noise. Arms, contexts, and trials were defined in the \textit{Background}. Settings for each of the parameters are listed in the following Subsection, \textit{Task Parameters}.

\textit{Reward Volatility} $\sigma$ is the standard deviation of the Gaussian random walk that governs the evolution of state-action reward probabilities throughout the task. Conversely, \textit{Reward Noise} indicates whether rewards in a given run included binomial noise. Noisy reward delivery refers to sampling of rewards from a Binomial distribution; conversely, noise-free reward delivery is the case in which a reward returned to the subject is equal in \textit{magnitude} to the probability that would otherwise parameterize the Binomial probability mass function. 

Within each condition, we simulated data from 30 artificial subjects whose parameters were generated according to procedures outlined in Subsection \ref{ss:model-params}. In sum, each condition (bound constraints vs. smooth constraints) was tested 8640 times (3 Arm levels, 4 context settings, 3 Trial length settings, 4 Reward SD levels, 2 Reward Noise conditions, with 30 synthetic subjects each). This design was selected in order to capture various possible scenarios in which models may be fit to behaviour collected from the K-context N-armed bandit class of tasks.

\subsection{Task Parameters}
\label{ss:tasks-params}

\begin{table}
    \centering
    \caption[Task parameters.]{Parameters of used tasks. \textit{Abbreviations:} K-Context N-Armed Bandit (NAB).}
	\label{tab:taskparams}
	\begin{tabular}{lc}
	\toprule
	\textbf{Parameters} 
	    & \textbf{KNAB}  \\ \midrule
	
	Number of Trials 
	    & $\lbrace 20, 100, 200 \rbrace$  \\ 
	
	Number of Arms
	    & $\lbrace 2, 4, 10 \rbrace$ \\ 

	Number of Contexts
	    & $\lbrace 1, 2, 5, 10 \rbrace$ \\ 
	    
	Reward Volatility
	    & \{0.01, 0.1, 0.5, 1\} \\
	    
	Reward Noise 
	    & \{ Noisy, Noise-Free \} \\

	\bottomrule
	\end{tabular}
\end{table}

The parameters of the simulations are presented in Table \ref{tab:taskparams}. The drifting reward probabilities (or magnitudes in the case of noise-free reward) were modelled according to a Gaussian random walk bounded by 0.2 and 0.8. At each time step, the previous reward probability is updated by adding random noise drawn from a standard normal distribution, scaled by $\sigma\sqrt{1/T}$. Within a given group of 30 simulated subjects, the same reward probabilities were used to ensure the 30 subjects differed from each other primarily in their model parameters.

\subsection{Models}
\label{ss:models}

For parsimony, we employed only a single model for this experiment. Parameters for subjects were generated along with the behavioural data; that is, no parameters were reused across conditions. 

All parameters were generated by sampling from uniform distributions over pre-defined closed intervals. The domains of support for each parameter were selected to be representative of a broad set of plausible values. Furthermore, to test our hypotheses about boundary effects, we ensured the parameters were drawn from the 98\% highest density interval (HDI) of the corresponding smooth prior probability distributions subsequently used as constraints. 

More specifically, the inverse softmax temperature was assumed to be Gamma(5,1) distributed, where the values and parentheses are the location and scale parameters, respectively. To generate the true parameters with which behavioural data were simulated, we sampled uniformly between the 1st and 99th percentiles of that distribution. As will be reviewed below, the smooth constraint used during model inversion (for the inverse softmax parameter) was a Gamma(5,1) probability density function. By this method, the true parameters were made to occupy a range that could be feasibly modeled by the smooth constraint condition, but yet demonstrated a uniform empirical distribution within those bounds. An exception to the above parameter generation procedure was implemented for the learning rate, which was sampled uniformly from the closed interval [0.2, 0.8].

\subsection{Model Fitting And Optimization}
\label{ss:optimization-method}

We composed a likelihood function for the simulated model, as per Equation \ref{eq:loglik-fx} (with softmax action selection and a Rescorla-Wagner learning rule). The arguments included the previously simulated actions, states, and rewards ($\mathbf{A}_i$, $\mathbf{S}_i$, $\mathbf{R}_i$) for a given subject, as well as current estimates of the respective subject's parameters $\mathbf{w}_i = \{\alpha_i, \beta_i\}$. A general template was provided in Algorithm \ref{alg:nab-nll-template}.

The parameters of the \texttt{patternsearch} and \texttt{fmincon} optimizers are presented in the appendix. For the former, a generalized pattern search with positive basis set cardinality of twice the input set dimensionality was chosen as the specific implementation. The greater density of search directions increases the probability of capturing a descent direction on the domain of the objective function \citep{Kolda2003}. Complete polling was used with the understanding that the number of function evaluations would increase substantially. However, this was a reasonable sacrifice owing to the present study's focus on parameter estimation accuracy and boundary effects of different constraints.

Optimization runs were nested in the innermost loop of the main experimental script, such that both constraint conditions were tested with the same input data and initial parameter estimates. This was done in an attempt to better control for factors other than the primary experimental condition. 

Initial values were sampled uniformly from the same range with which the ground truth parameters were sampled. This was done for the purpose of eliminating poor initialization as an explanation of any differences observed. For instance, initialization outside of the feasible region may affect smooth vs. bound constraints differently. Smooth constraints may demonstrate some descent in the objective function if initialized in a region within the prior probability density's domain of support, regardless of whether it lies outside of our feasible region. Conversely, initialization outside of the feasible region when bound constraints are present could potentially lead to projection of that point on to the bound constraint plane. If the bound constraint plains are indeed "sticky," then this projection onto the boundary would result in the hypothesized truncation effects. However, we are not simply interested in stickness of the bound constraint, but also whether they are attractive to nearby trial iterates. 

We thus wanted to minimize the effects of projection onto the constraint boundary, and instead test whether optimizers initialized in a range containing the true underlying value would gravitate toward the bounds. As such, all runs were initialized within the feasible zone. In case the optimizers became stuck at local minima, we allowed up to 100 restarts with new initializations. We also did this to evaluate whether the constraint planes induced spurious local minima that could be overcome by restarting. The number of function evaluations was summed over all restarts to return the final (total) number of function evaluations required to optimize the likelihood function. The algorithm was restarted with a new initialization until no further improvement in objective function value was produced (at which point the estimates computed at the previous run were taken as the final values). 

The constraints passed to the optimizers for each parameter are shown in Table \ref{tab:model-param-constraints}. Note that while we generated learning rates by sampling from a uniform distribution spanning the interval [0.2, 0.8], the constraints passed to the optimizers for this parameter were [0, 1]. Setting the bounds for learning rate estimation to [0, 1] kept the bound constraint scenario more comparable with the smooth constraint condition, since the Beta distribution is supported on (0, 1). It also provides a clear scenario in which all true underlying parameters are far from the constraint planes. This eliminates the possibility that any final estimates "sticking" to the bound constraints would be due to the true values lying nearby. 

\begin{table}
    \centering
    \caption[Constraints]{Constraints passed to the optimizers for each parameter. For instance, all inverse softmax temperatures were given the same constraints, regardless of whether they were situated in the lrcr, lrcrp or other models. \textit{Abbreviations and Symbols:} learning rate ($\alpha$), inverse softmax temperature ($\beta$). Note the bound constraints on the inverse softmax temperature $\beta$ represented the 1st and 99th percentile values of Gamma(5, 1). }
	\label{tab:model-param-constraints}
	\begin{tabular}{lcc}
	\toprule
    \textbf{Parameter} 
	    & \textbf{Bounds [Low, High]}
	    & \textbf{Smooth}\\ \midrule
	
	% NAB MODELS ------------------------
	$\alpha$
	    & $[0, 1]$
	    & Beta(1.1, 1.1)\\
	
	$\beta$
	    & $[\approx 1.28, \approx 11.6]$
	    & Gamma(5, 1) \\
	\bottomrule
	\end{tabular}
\end{table}

\subsection{Statistical Analysis}
\label{ss:statistical-analyses}

The first outcome measure in the present study was the $\ell_2$ norm of the difference between the (subject-level) estimated parameters $\hat{\bold w}_i$ and the ground truth parameters generated in simulation $\boldsymbol\theta_i$. Learning rate and choice randomness parameters were rescaled onto a common range (otherwise the inverse softmax temperature would dominate the $\ell_2$ norm). We then used the $\log \ell_2$ norm for each subject-level estimate as the response variable in a Bayesian linear regression model against constraint type (bound vs. smooth), optimizer (\texttt{patternsearch} vs. \texttt{fmincon}), and an interaction between the two. We confirmed by visual inspection that the $\log \ell_2$ norm was approximately Gaussian in distribution. Covariates included the number of trials, number of contexts, number of arms, reward noise, and reward volatility. A Gaussian prior with mean 0 and unit variance was set over the linear regression weights (and bias term). The standard deviation on the Gaussian over the $\log \ell_2$ norm was set at 1. We fit this model using mean-field variational inference in the Edward package \citep{Tran2016} in Python. The importance of each variable was assessed by sampling 1000 instances of the corresponding linear regression parameter from the approximate posterior distribution and determining whether the 95\% HDI overlapped with 0. 

The second outcome measure concerned the form of the quantile-quantile plots generated by the bound vs. smooth constraint conditions in comparison to the ground truth condition. Specifically, our hypothesis regarding truncation effects with boundary constraints would manifest as sharp-cornered, flat regions at the extremes of the Q-Q plot. The corresponding regions for the ground truth parameters and smooth constraint estimates would conversely look smooth (reflecting preservation of relative order in parameter estimate magnitudes). 

In order to investigate optimizer behaviour in relation to boundary effects, we plotted a convergence graph of $\log\{Step\,Size\}$ over $FunctionEvaluations$ for runs in which the final result was stuck to the bound constraint, and for runs that did not show this behaviour. We classified runs as "stuck" to the bounds as follows. First, we took the absolute difference between the parameter estimate (both learning rate and inverse softmax temperature) and the respective bounds in Table \ref{tab:model-param-constraints}. These absolute differences was then scaled between 0 and 1 for each parameter so that both were on the same scale. The result is that estimates that were furthest from the bounds were those exactly between the two boundaries (at least in the case of learning rate, which in both constraint conditions was between 0 and 1). Stuck runs were then defined as those where the scaled absolute difference was less than 0.01 for either parameter. 

\subsection{Implementation}
\label{ss:implementation}

All experiments were implemented in Matlab R2017a (The MathWorks; Natick, MA). Statistical analyses were conducted in the Python programming language using the Edward package \citep{Tran2016} and custom scripts.

\section{Results}
\label{s:results}

Results from the linear regression of the $\log \ell_2$-norm of the parameter estimate error are listed in Table \ref{tab:lr-results}. The 95\% highest density interval of the approximate posterior over linear regression weights included zero for the optimizer variable (median of -0.004, 95\% HDI -0.019 to 0.017), but was above zero for the bound constraint (median of 0.063 , 95\% HDI of 0.045 to 0.078). This suggests that bound constraints induced a slight worsening of parameter estimate accuracy when controlled for the number of trials, contexts, arms, reward noise, and reward volatility.

\begin{table}
\centering
\caption{Results of linear regression of $\log \ell_2$-norm of the difference between ground truth and estimated parameters. \textit{Abbreviations:} highest-density interval (HDI), pattern search (PS).}
\label{tab:lr-results}
\begin{tabular}{lcc}
\toprule
\textbf{Variable} & \textbf{Median} & \textbf{95\% HDI} \\
\midrule

Intercept                              & -0.636 & (-0.646, -0.626)\\
Trials                                 & -0.439 & (-0.451, -0.429)\\
Contexts                               &  0.072 & (0.051,0.089)   \\
Arms                                   & -0.163 & (-0.175,-0.153) \\
Reward Noise                           &  0.074 & (0.065,0.084)   \\
Reward Volatility                      & -0.034 & (-0.043,-0.022) \\
Optimizer == PS                        & -0.004 & (-0.019,0.017)  \\
Constraint == Bound                    &  0.063 & (0.045,0.078)   \\
Constraint == Bound \& Optimizer == PS & -0.007 & (-0.023,0.009)  \\
\bottomrule
\end{tabular}
\end{table}

\begin{figure*}
\centering
\includegraphics[width=\textwidth]{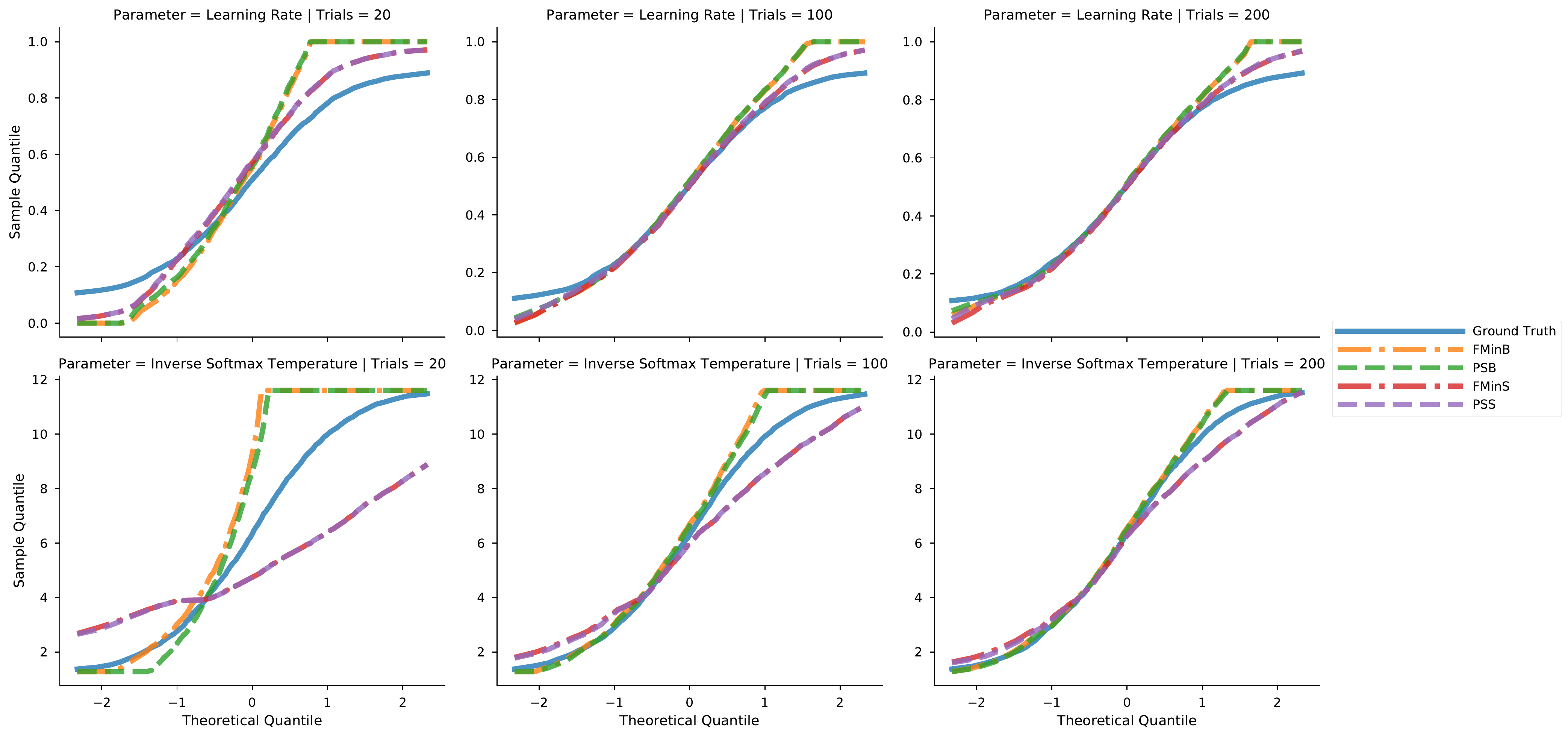}
\caption{Quantile-quantile plots for ground truth (in blue) and estimated parameters faceted by parameter (rows) and number of trials (columns). Each line represents a different optimizer/constraint combination. \textit{Abbreviations:} \texttt{fmincon} with bound constraints (FMinB), \texttt{fmincon} with smooth constraint (FMinS), pattern search with bound constraint (PSB) and pattern search with smooth constraint (PSS). Note the truncation in models with bound constraints, while those using smooth constraints retain a nonzero slope.}
\label{fig:qq-plots}
\end{figure*}

Quantile-quantile (Q-Q) plots of the ground truth and estimated model parameters are presented in Figure \ref{fig:qq-plots}. One can appreciate the substantial truncation effects observed with the use of boundary constraints across all parameters estimated. This observation was not limited to estimates of the inverse softmax temperature, where the bounds given to the pattern search optimizer were equal to the bounds on the true parameters (which are shown as the red curves in the Q-Q plots). Recall that for learning rates, we sampled the ground truth parameters from the closed interval [0.2, 0.8], but supplied pattern search with bound constraints of [0, 1]. Despite the true parameters falling well within the optimization bounds, one continues to appreciate substantial truncation effects around the boundaries. These effects were not present with the use of smooth constraints, with which one can appreciate the relative preservation of ordinal relationships between parameter estimates.

Of note, the beneficial effect of trial sample size is apparent in Figure \ref{fig:qq-plots}, where the degree of truncation (as measured by the length of the flat regions at the extremes) becomes progressively shorter with more trials. 

\begin{figure*}
\centering
\includegraphics[width=\textwidth]{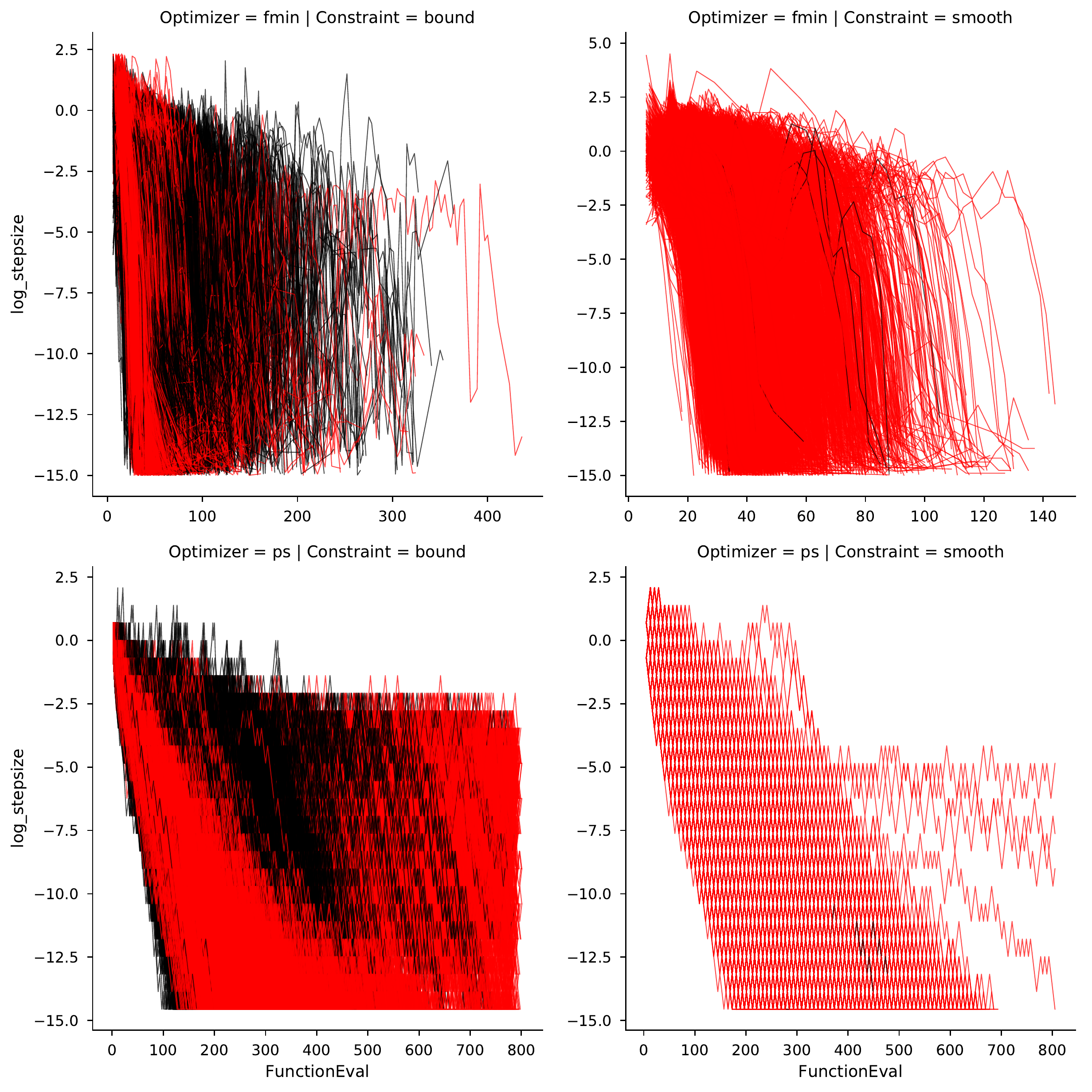}
\caption{Convergence graph plotting $\log\{Step\,Size\}$ (y-axis) against cumulative function evaluations (x-axis) for the \texttt{fmincon} optimizer (top row) and \texttt{patternsearch} optimizer (bottom row). Bound constraint conditions are shown in the first column of plots, and smooth constraint conditions are shown in the second column. Each line is a single run of an optimizer on the behavioural data. Red lines are those runs that did not lead to final parameter estimates at the bound constraints, Black lines are those runs that stuck to bound constraints at optimizer termination. }
\label{fig:convergence-plot}
\end{figure*}

Figure \ref{fig:convergence-plot} plots the $\log\{Step\,Size\}$ against cumulative number of function evaluations for each optimizer $\times$ constraint combination. It is notable that smooth constraints were generally associated with more rapid convergence for both \texttt{fmincon} and \texttt{patternsearch}. Moreover, pattern search under smooth constraints shows linear convergence (i.e., reduction in step size by a roughly constant factor at each function evaluation), whereas a much more variable pattern of convergence can be observed under bound constraints for that method. Under \texttt{fmincon}, one observes a greater variability in step size and duration of the search under the bound constraint condition; that is, runs that did not stick to the boundary appear to converge rapidly, whereas a much larger proportion of those runs that extend beyond 150 function evaluations resulted in truncation effects.

\section{Discussion}
\label{s:discussion}

For a canonical RL task, we have shown that fitting computational cognitive models using optimization procedures with bound constraints can result in substantial truncation effects around the constraint boundaries, and slightly worse parameter estimation accuracy. Use of boundary constraints appears to bias estimates toward lying along the bounds themselves. This effect was overcome slightly by increasing the number of behavioural trials in an experimental run (of the RL task), but was not abolished with up to 200 trials.

Our findings are especially relevant for studies in which between-individual and between-group comparisons are important. In these studies, the ordering of parameter estimate values may be arguably more important than the values themselves. For instance, in determining the propensity of depressives to undervalue rewards, it may suffice to evaluate whether their reward sensitivity parameters are lower than those of healthy subjects. In other words, we would aim for such phenotypic relationships to be monotonically invariant between experiments. Indeed, if comparing such results across different task paradigms, the actual values of parameter estimates may be less important than this ordering. The use of bound constraints during estimation of these parameters may thereby hinder such comparisons, since truncation effects destroy any ordinal relations between individual parameter estimates near the boundaries. Conversely, the use of smooth constraints such as prior probability density functions could preserve those ordinal relationships. 

\subsection{Whence the Truncation?}

The boundary effects observed in the comparisons between smooth and bound constraints are necessarily a product of the optimization, since both conditions were implemented on the same data at each run. For the \texttt{fmincon} method, a trial iterate is located within the feasible region prior to evaluation of its candidacy as the subsequent incumbent. The \texttt{patternsearch} implementation of bound constraints is slightly different. The function chain in Matlab proceeds as follows (beginning with \texttt{patternsearch}, and where the functions lower in this list called by those immediately preceding them):

\begin{enumerate}
\item \texttt{patternsearch}
\item \texttt{pfminbnd}
\item \texttt{poll}
\item \texttt{boxdirections}
\end{enumerate}

\noindent The \texttt{pfminbnd} is a function private to pattern search. The deepest function, \texttt{boxdirections} returns a positive basis to be used for search directions. That function first determines whether bound constraints are active at the current point before constructing a tangent cone that specifies the space that our search directions must positively span. When an adaptive mesh is used (which is by default in \texttt{patternsearch}), the tangent cone is limited by a set of vectors parallel to the active constraints. In the \texttt{poll} function, the tangent cone is added to the set of direction vectors. Thus, in the coming iteration(s), the candidate solutions evaluated by pattern search will include those lying on or close to the boundary. 

The present study collected step-size output from the optimizers in order to evaluate whether any systematic changes in step size were evident between conditions. We had initially suspected that the optimizers might have experienced a rapid decline in step size upon hitting the boundary, but this was not observed. Indeed, there was no systematic growth or shrinkage of the step size upon hitting the boundary constraint. Rather, there appeared to be much greater volatility in step size---in addition to a need for more function evaluations---in conditions where the optimizer (particularly \texttt{fmincon}) stuck to the constraint plane. Boundary constraints may thus change the shape of the objective function (from the perspective of the optimizer) in a fashion that precludes smooth adaptation of step sizes at each iteration. However, since repeated restarts did not eliminate this effect, we believe it is unlikely that the bound constraint approach merely induced local minima.

\subsection{Strengths and Limitations}
\label{ss:limitations}
The primary strength of the present study was the development of a testbed for a task class commonly used in the computational psychiatry literature, which was analyzed in a fashion capturing a broad set of potential use cases in human studies. By simulating this wide set of conditions, we were able to assess the likely effects of using bound constraints in practice. Moreover, our methods permitted a direct comparison of bound and smooth constraint conditions, because every optimizer/constraint condition was applied to the same simulated data.

One limitation of the present study includes the relatively few prior probability density constraints implemented. Specifically, we did not investigate the relationship between the location of the prior probability density and the resulting estimates. However, with larger trial numbers, the location of the prior probability density becomes less important, since the likelihood (i.e., the subject's data) eventually dominates the objective function value. In contrast, bound constraints are similarly enforced regardless of sample size. As such, we do not believe that the results of the present study are substantially impacted by this choice.

Another limitation is the use of only one model, which we did for parsimony. However, we also chose this model on account of it representing the relatively simplest non-trivial case. Indeed, we showed that in this base scenario, bound constraints may distort parameter estimation. It is reasonable to suspect that models with greater complexity would be similarly, if not more, affected by the truncation of bound constraints. That being said, further experimentation would be warranted in future work.

\subsection{Conclusions}
\label{ss:conclusions}

In sum, we have shown that implementing smooth penalty functions---in this case a prior probability density over free parameters---can avoid truncation effects when fitting reinforcement learning models to behavioural data. These effects may be caused by readjustment of search directions when the search reaches a constraint boundary, and may be partially overcome by increasing the number of behavioural trials sampled from subjects. Our findings may have methodological implications for studies in computational psychiatry and cognitive neuroscience, but should be further examined using other optimization algorithms and RL models.

\vspace{\fill}\pagebreak

% ITEM 9 [See the "howto.tex" file.]
\appendix
\renewcommand{\theequation}{A\arabic{equation}}
\setcounter{equation}{0}
\renewcommand{\thesection}{\Alph{subsection}}
\setcounter{section}{0}
\section{Appendix A: Optimizer Options}
\label{app:ps-opts}

Options passed to the Matlab \texttt{patternsearch} function are presented in Table \ref{tab:psoptimset}. For the \texttt{fmincon} optimizer, we set "TolFun"=1e-14, "TolX"=1e-14, "Display"='off', and the remainder of parameters were left as defaults.

\begin{table*}[!htpb]
\caption{Settings used for the Matlab \texttt{patternsearch} function. To implement these options in Matlab, one would specify \texttt{opts = psoptimset('Keyword1', 'Argument1', 'Keyword2', 'Argument2', ...);} as per the pairs specified below. Sets of options within curly braces $\lbrace \rbrace$ identify the settings that were varied in the present experiment. \textit{Abbreviations:} generalized pattern search (GPS), input space dimensionality (\texttt{n$\_$dim} or simply $n$). Options not listed here were set to their default values.}
\label{tab:psoptimset}
\begin{tabular}{p{0.3\textwidth} p{0.3\textwidth} p{0.3\textwidth}}
    \textbf{Keyword} & \textbf{Argument} & \textbf{Remark} \\ \hline
        'PollMethod' 
            & 'gpspositivebasis2n'
            & \\

        'CompletePoll' 
            & 'on'
            & All $2n$ points were evaluated at each iteration. \\
            
        'PollingOrder' 
            & 'Consecutive'
            &  \\
        
        'CompleteSearch'
            & 'off' 
            & Search step not implemented. \\
        
        'SearchMethod' 
            & []
            & Search step not implemented. \\

        'MeshAccelerator'
            & 'off'
            & \\

        'ScaleMesh'
            & 'on'
            & \\

        'MeshExpansion'
            & 2
            & \\
        
        'MeshContraction'
            & 0.5
            & \\
        
        'TolMesh'
            & 1e-14
            & \\
        
        'TolX'
            & $1e^{-14}$
            & \\
        
        'TolFun'
            & $1e^{-14}$
            & \\

        \hline
\end{tabular}
\end{table*}
\vspace{\fill}\pagebreak

% ITEM 10 [See the "howto.tex" file.]
\bibliographystyle{plainnat}
\bibliography{bibliography}

\end{document}